%Paper: cond-mat/9408096
%From: jpr@viking.lanl.gov (Jose P. Rodriquez)
%Date: Wed, 31 Aug 94 02:16:36 MDT

%Tex file.
\def\gsim{\mathrel{\scriptstyle{\buildrel > \over \sim}}}
\def\lsim{\mathrel{\scriptstyle{\buildrel < \over \sim}}}
%\magnification=1095
\magnification 1200
\baselineskip=17pt
%\font\twelverm=cmr10 scaled 1200

%\vskip 32pt
\centerline{\bf OBSERVABLE ZERO-SOUND IN STRONGLY CORRELATED METALS}
\bigskip
%\centerline{\bf ELECTRONS IN THE OVERDOPED LIMIT}
\vskip 50pt
\centerline{J. P. Rodriguez}
\medskip
\centerline{Theoretical Division,
Los Alamos National Laboratory,
Los Alamos, NM 87545.\footnote*
{Permanent address: Dept. of Physics and Astronomy,
California State University,
Los Angeles, CA 90032.}}
\vskip 30pt
\centerline  {\bf  Abstract}
\vskip 8pt\noindent
The slow zero-sound mode expected near the Mott transition
in strongly
interacting two-dimensional
Fermi systems that are neutral
 is shown to persist as the physical
sound mode in the case that
the fermion carries electronic charge and is
embedded in a positive ionic background.  The latter
sound velocity softens completely precisely at
the Mott transition, indicating that a zone-center
structural transition will occur in the system.
We suggest that this phonon-softening mechanism
is related to the structural
transitions commonly observed in the oxide
superconductors.
\bigskip
\noindent
PACS Indices:   71.27.+a, 71.45, 63.20.K, 74.20.Mn
\vfill\eject

The realization that the conduction electrons
in the Copper-Oxygen planes common to all
high-temperature superconductors
experience strong repulsive interactions
has motivated the ongoing theoretical study
 of strongly interacting fermions in two dimensions.$^1$
The simplest theoretical description of this physics for
electron densities approaching  the antiferromagnetic
Mott transition is given by the $t-J$ model on the
 square-lattice,$^{2,3}$ where double occupancy of
electrons at each site is strictly excluded.  The
latter constraint can be elegantly accounted for
 by the introduction of an auxiliary slave-boson
field, $b_i$, such that the original electron field,
$c_{i\sigma}$, is replaced by the composite field,
$c_{i\sigma}b_i^{\dag}$, along with the new constraint,
$$\sum_{\sigma}c_{i\sigma}^{\dag}c_{i\sigma}+
b_i^{\dag}b_i=1.\eqno (1)$$
Certain mean-field treatments
based on this formulation
of the $t-J$ model that allow for dynamics in the
slave-boson field lead to spin/charge
separation of the strongly correlated electron.
As a result, the latter break-up of electronic
quantum numbers manages to reconcile
the experimental observation of an electron-type
Fermi surface in conjunction with hole-type charge
transport
in the normal state of the oxide superconductors.$^{2-5}$
Such mean-field spin/charge separated treatments,
however, over-estimate  the size of the Hilbert space
allowed by the constraint against double occupancy (1);$^6$
e.g., the constraint against double occupancy at a particular
site is violated.

Beyond the mean-field approximation, however,
there exist fluctuations
of the statistical gauge field, $A_{\mu}$,
%that is
generated by the internal
gauge symmetry $(c_{i\sigma}, b_i)\rightarrow
(e^{i\theta_i}c_{i\sigma}, e^{i\theta_i}b_i)$
that mediate interactions between the
 spin/charge separated  species.
In particular, slave-boson
scattering off of such gapless
{\it transverse} gauge-field excitations
results ultimately
in a linear-in-$T$ prediction for resistivity in the fluxless
(``strange'') metallic phase of the $t-J$ model that
is  consistent with transport measurements
in the normal phase of the oxide superconductors.$^4$
It has  been pointed out by the author, however,
that a gapless pole is also shown by the {\it longitudinal}
component of the statistical gauge-field, which is directly
tied to the constraint against double occupancy (1).$^5$
Physically, this mode is a slow
zero-sound mode with a velocity given by $c_0\sim (tJx)^{1/2}$,
where $x$ denotes the concentration of holes.$^5$
(Throughout this paper, we set $\hbar$, $k_B$,
and the lattice constant, $a$, to unity.)
Unlike the case of
a conventional Fermi-liquid, where sound excitations
that are slower than the Fermi velocity become over-damped
by the particle-hole continuum,$^7$ the slow zero-sound mode
in question remains under-damped because of spin/charge
separation.$^8$
Recently, Hlubina et al.
have shown that including
such longitudinal gauge-field fluctuations, in general,
results in an entropy reduction.$^6$  In fact,
 they found that the addition of the spin/charge
separated quasi-particle entropy with the entropy
``generated'' by  the gauge-fields
agrees quantitatively with a calculation
of the entropy for the $t-J$ model on the
square-lattice that is based on  a
high-temperature expansion.$^6$
Therefore, the introduction of the longitudinal
gauge-field cures the Hilbert space size problem
existing in the mean-field approximation mentioned
above, thereby demonstrating that the presently
discussed gauge-field theory  is a
consistent low-energy effective theory
of strongly interacting electrons in two dimensions.
Note that a very similar slow zero-sound
mode also appears in the anyon superconductor saddle-point
of the $t-J$ model near half-filling.$^9$

In this Letter, we show that the above-cited slow
zero-sound mode existing
in neutral two-dimensional (2D)
Fermi systems near the Mott transition$^{5,9}$ persists
when such fermions are promoted to physical
electrons embedded in
a positive ionic background.  In particular, we compute
the sound velocity for a strongly interacting electron
liquid embedded in a featureless jellium background
and find that it interpolates continuously between the standard
Bohm-Staver result for first sound at
hole densities far from the Mott transition$^{10}$
and the slow zero-sound velocity at hole
densities approaching the Mott transition
($x\rightarrow 0$).  We therefore argue that a zone-center
structural transition will be induced at the Mott transition,
where the sound speed softens completely.   We also
compute the renormalized velocity of acoustic phonons
coupled
to the same strongly interacting electron liquid via the
standard electron-phonon interaction and find very
similar results.  (It is implicitly assumed in the latter
calculation that the long-range Coulomb interaction is
screened by some other type of free charge carriers in the system.)

{\it Zero-sound + Jellium.}  The longitudinal dielectric constant
for any electronic
liquid embedded in a positive jellium background may be
expressed as
$$\epsilon(\vec k,\omega)=1-{\Omega_i^2\over{\omega^2}}
+{\Omega_e^2\over{c_0^2k^2(1+i\nu_1\rho_F{\omega\over k})^{-1}
-\omega^2}}\eqno (2)$$
in the long wavelength limit, where $\Omega_e$ and
$\Omega_i$ denote the
plasma frequencies of the electron liquid and of
the positive jellium background, respectively,
and where the zero-sound velocity, $c_0$,
is related to the Fermi-Thomas wave-vector
of the electron liquid, $k_{FT}=(4\pi e^2\rho_F)^{1/2}$, by
$$\Omega_e=c_0 k_{FT}.\eqno (3)$$
Above, $\rho_F$ represents the density of states
at the (pseudo) Fermi surface.
Also, the  prefactor, $\nu_1$, to the Landau damping
term in eq. (2) is of order unity for $|\omega/k|< v_F$,
where $v_F$ denotes the (pseudo) Fermi velocity,
while it vanishes otherwise.
In the case of the spin/charge separated ``strange''
metal phase of the square-lattice $t-J$ model
at hole concentrations near half-filling
($x\rightarrow 0$),$^5$
the electronic plasma frequency is given by
that of the holes, $\Omega_e^2=(32/\pi) (e^2/d) tx$,
while the density of states at the pseudo Fermi
surface is given by $\rho_F\gsim (d J)^{-1}$.
(We presume that an infinite stack of such
$t-J$ models,
each separated by a distance $d$, fills three-dimensional space.)
Hence, eq. (3) implies that the zero-sound
velocity is approximately $c_0\sim (tJx)^{1/2}$
in this case,$^5$ as we quoted earlier.
The spectrum for the collective charge-modes
of this system is determined by
the characteristic equation $\epsilon(\vec k,\omega)=0$,
which yields the following two solutions:
$$\eqalignno{\omega_-(\vec k)=
&c_s k\Bigl(1+{i\over 2}\nu_1\rho_Fc_s\Bigr),
&(4a)\cr
\omega_+(\vec k)=
&\Biggl[\Omega_e^2+\Omega_i^2+c_0^2
\Biggl(1+{c_1^2\over{c_0^2}}\Biggr)^{-1}k^2\Biggr]^{1/2},
&(4b)\cr}$$
where the sound speed is given by
$$c_s=(c_0^{-2}+c_1^{-2})^{-1/2}, \eqno (5)$$
and where the bare first-sound speed, $c_1$, is related to the
plasma frequency of the
%jellium
ionic background by$^{10}$
$$\Omega_i=c_1 k_{FT}.\eqno (6)$$
In deriving the dispersion for the sound mode (4a),
it has been assumed that the Landau damping is a small perturbation,
$\nu_1\rho_Fc_s\ll 1$, which in the case
of the ``strange'' metal phase of the $t-J$ model implies that
$tx\ll J$.$^5$

Above, we have obtained the acoustic sound mode (4a) and the
plasma oscillation (4b) generally expected in a two-component
plasma.$^{11}$  In the case of a conventional metal, where the first-sound
speed is much smaller than the zero-sound speed ($c_1\ll c_0$),
we recover the well-known result that the zero-sound mode
is ``pushed-up'' to the plasma frequency,$^{7}$ while that the sound
speed is given by the Bohm-Staver result,$^{10}$ $c_1$.
(Note that {\it transverse} zero-sound is predicted to exist
both in charged Fermi-liquids$^{12}$ and in conventional
metals.$^{13}$)
On the other hand, for hole-densities approaching the
Mott transition such that $\Omega_e\ll\Omega_i$,
the sound speed (5) is then approximately
that of zero-sound; i.e., $c_s\cong c_0$.  Hence,
precisely at the Mott transition, where $\Omega_e=0$ and $c_0=0$,
Eq. (4a) indicates that the acoustic phonon mode softens completely,
suggesting that a zone-center structural transition will occur
in the system.

{\it Zero-sound + Phonons.}$^5$  Consider now the  case of
the very same strongly interacting electron liquid
with a dynamic compressibility [see the last term in eq. (2)]
given by
$$\kappa(\vec k,\omega)=
\rho_F\Biggl[\Bigl(1+i\nu_1\rho_F{\omega\over k}\Bigr)^{-1}
-{\omega^2\over{c_0^2k^2}}\Biggr]^{-1}\eqno (6)$$
that couples to acoustic phonons via the conventional electron-phonon
interaction.  Then standard diagrammatic techniques based
on Dyson sums yield that the renormalized phonon propagator,
$D(\vec k,\omega)$,
satisfies
$$D^{-1}=D_0^{-1}+g^2\kappa, \eqno (7)$$
where $D_0(\vec k,\omega)=c_1^2k^2(\omega^2-c_1^2k^2)^{-1}$
is the bare
phonon propagator and $g$ denotes the coupling-constant
for the electron-phonon
interaction.$^{14}$  Substituting (6) into (7), we find that the
renormalized phonon propagator, $D(\vec k,\omega)$, has poles
of the form
$[\omega-c_{\pm}k-i\gamma_{\pm}(\vec k)]^{-1}$, with
phonon velocities and damping rates given by
$$\eqalignno{c_{\pm}=&
\Biggl\{ {1\over 2}(c_0^2+c_1^2)\pm
{1\over 2}[(c_0^2-c_1^2)^2+4\rho_Fg^2c_0^2c_1^2]^{1/2}\Biggr\}^{1/2},
& (8a)\cr
\gamma_{\pm}(\vec k)=&
{1\over 2}\nu_1\rho_F c_0^2 k\Biggl
\{{1\over 2}\pm{1\over 2}{c_0^2-c_1^2\over
{[(c_0^2-c_1^2)^2+4\rho_Fg^2c_0^2c_1^2]^{1/2}}}\Biggr\}.
& (8b)\cr}$$
In deriving the dispersion for the phonon modes (8a,b),
it has been assumed that the Landau damping is a small perturbation;
i.e., $\gamma_{\pm}(\vec k)\ll c_{\pm}k$.
This again requires that $\nu_1\rho_Fc_s\ll 1$, which implies
that $tx\ll J$
for the case of the ``strange'' metal phase of the $t-J$ model.$^5$

We see above in eqs. (8a) and (8b) that the coupled
electron-phonon system
results in two acoustic modes that are weakly damped.
As shown in Fig. 1,
the two branches of velocity exhibit  level-repulsion
characteristic of
reactively coupled modes.  In the conventional metallic regime,
where $c_1\ll c_0$, while we do
recover the well-known result
$c_-\cong c_1(1-\rho_Fg^2)^{1/2}$ for the
renormalized phonon velocity,$^{14}$
we also obtain an observable zero-sound branch
with velocity $c_+\cong c_0$.  The latter result
appears to contradict our previous calculation in the jellium system,
which found that the zero-sound mode was promoted to
the plasma frequency in this regime.  However, it can be shown
that the presently obtained acoustic modes can be
recovered in the former
jellium model as long as the vacuum dielectric constant,
$\epsilon_0=1$, in Eq. (2)
is replaced by one with static screening;$^{15}$ i.e.,
$\epsilon_0=1+k_{FT}^{\prime 2}/k^2$.
Such static screening could
result from the existence of some other type of charge-carriers
in the system.$^{5}$  On the other hand,
in the limit approaching the Mott transition,
where $c_0\ll c_1$, eq. (8a) indicates that the
renormalized phonon velocity, $c_-$,  evolves continuously
into the zero-sound velocity, $c_0$ (see Fig. 1).
Hence, just as in
the previous case of the jellium model, this
phonon will soften completely at the Mott transition,
$c_0=0$, suggesting that a zone-center structural
transition will occur in the system.  Last,
we take this opportunity to point out
that the preceding result corrects previous
work by the author,$^5$ where it was erroneously
asserted that complete phonon softening occurs
when $c_0=c_1$.

{\it Observability.}  We have shown above that slow zero-sound evolves
continuously into the first-sound mode in
strongly correlated metals near the Mott transition.  The conditions for
the observability of such a mode are therefore
({\it a}) that it exist in the neutral system,
and ({\it b}) that the plasma frequencies satisfy
$\Omega_e\lsim\Omega_i$.  For the case of
the spin/charge separated ``strange'' metal phase of
the $t-J$ model in two dimensions,$^{4}$  the
former requirement  reduces
to the acoustic-plasmon type existence condition,$^{5,11}$
$$v_B\ll c_0\ll v_F,\eqno (9)$$
where $v_B$ denotes the characteristic velocity of the
slave-bosons (holons), and where $v_F$ denotes the
characteristic velocity of the pseudo-fermions (spinons).
 Near half-filling, since $v_F\sim J$ and $c_0\sim (tJx)^{1/2}$,
 the righthand side of eq. (9)
implies ({\it i}) that $tx\ll J$.
Now suppose first that no
superfluid transition exists in the spinon component.  It can
then be shown that the holon Bose-condensation transition$^{16}$
is suppressed to zero-temperature by chiral
spin-fluctuations,$^{17,18}$
which implies that the
characteristic velocity of the holon liquid
is given by the thermal velocity, $v_B\sim (tT)^{1/2}$,
at all temperatures $T$.  Hence, the lefthand
side of eq. (9) implies ({\it ii}) that
$T\ll Jx$, in this case.  These
results indicate, therefore,
that slow zero-sound exists in strongly interacting
2D Fermi
systems  near the Mott transition and  at
low-temperature when
Cooper pairing instabilities in the spinon fluid
are suppressed.
If, on the other hand, we suppose that there does
indeed exist a superfluid transition in the spinon
component at a critical temperature, $T_F$,
 then the critical temperature
of the physical superconducting transition will be given
by the geometric mean $T_c=(T_F^{-1}+T_B^{-1})^{-1}<{\rm min}(T_F,T_B)$,
where $T_B\sim tx$ is the Bose-Einstein condensation temperature
of ideal holons.$^{18}$  In such case, the  characteristic velocity
of the holons is given by $v_B\sim (tT_c)^{1/2}$ in the
 superconducting phase, and the lefthand
side of eq. (9)  then implies ({\it iii}) that  $T_c\ll Jx$.
However, for $t>J$, the latter inequality  can only be satisfied
at hole concentrations {\it beyond} that which optimizes
$T_c$, since $T_c\cong T_B\sim tx$ for
small hole concentrations, $x\rightarrow 0$.
In the context of the oxide superconductors, therefore,
the above discussion  suggests that slow zero-sound of the type
sketched here is observable in the optimally-doped to
over-doped regions of
the phase diagram, where spin-gap behavior in the normal
state is absent.$^{16-19}$

It is interesting to remark that structural
transitions in the cuprate superconductors are known
to occur precisely in the above regime.$^{20}$
This observation should be tempered, however, by the fact that the
oxide superconductors appear to be far from a Mott transition
at such hole concentrations $(x\gsim 0.1)$.  In particular, the
condition ({\it b}) $\Omega_e^2\lsim c_1^2 k_{FT}^2$
for proximity to the Mott transition yields an upper bound of
$x\lsim (c_1/ta)^2(2t/J)
\sim 0.01$ for the hole concentration in this case,$^5$
where we have taken
$\Omega_e^2\sim 10 tx (e^2/d)$, along with model parameters
$t/J\sim 5$, $t=0.5\, {\rm eV}$, $a=4 \, {\rm \AA}$, and a first sound
speed $c_1\sim 10\,{\rm km/s}$ characteristic of such materials.$^{21}$
Mean-field treatments of the square-lattice $t-J$ model
find,$^9$ however, that the metallic saddle-points  become unstable
to a Mott insulator at $tx\lsim 0.02 J$.  Hence,
slow zero-sound should be observable within
a window of  mobile hole concentrations
satisfying $0.001 \lsim x \lsim 0.01$ in the oxide superconductors.

In summary, the main result established in
this paper is that the slow zero-sound mode that appears
in certain spin/charge separated treatments of
the problem of strongly interacting electrons
in two dimensions  near the Mott transition$^{5,9}$ is observable,
in the
presence of an ionic background, as the physical
sound mode for hole concentrations that satisfy
$\Omega_e\ll\Omega_i$.
On the other hand,  far from the Mott
transition in the conventional metallic regime,
$\Omega_e\gg\Omega_i$, it evolves continuously
into the standard Bohm-Staver first-sound mode.$^{10}$
The former result also implies that the
sound speed softens {\it completely} precisely at the
Mott transition, $\Omega_e=0$, which suggests that a
zone-center structural transition will occur
in the system.
It is useful to compare the
present mechanism for electronically driven zone-center
structural transitions with
those based on strong electron-phonon coupling,$^{14}$
$\rho_Fg^2=1$, and with mechanisms for
zone-edge structural transitions
that rely on the appearance of a Peierls distortion
in the system.$^{22}$
Last, since the present results rely essentially
on the collapse of the Drude weight near the Mott
transition, they could well be valid
for any strongly interacting
Fermi system in two dimensions.
Such is the case for example in one dimension, where the
velocities for collective charge excitations
in the Hubbard model and in the $t-J$ model
vanish at half-filling in an identical fashion.$^{8}$

The author thanks N. Schopohl, A.J. Leggett,
P. Lederer, A. Migliori and S. Billinge
for discussions, and especially
A.J. Millis
for helping uncover the error in previous work that
was referred to earlier.  The present work
was performed under the auspices of the
Department of Energy, and  was supported
in part by the Associated Western Universities
and by National Science Foundation grant
DMR-9322427.

\vfill\eject
\centerline{\bf References}
\vskip 16 pt

\item {1.}  P.W. Anderson, Science
 {\bf 235}, 1196 (1987).

\item {2.} L.B. Ioffe and A.I. Larkin,
 Phys. Rev. B {\bf 39}, 8988 (1989).

\item {3.} J.P. Rodriguez and
 B. Dou{\c c}ot, Europhys. Lett.
{\bf 11}, 451 (1990).

\item {4.} N. Nagaosa and P.A. Lee,
Phys. Rev. Lett. {\bf 64},
2450, (1990);  L.B. Ioffe and
 P.B. Wiegmann, Phys. Rev. Lett. {\bf 65},
653 (1990); L.B. Ioffe and
 G. Kotliar, Phys. Rev. B {\bf 42}, 10348 (1990).

\item {5.} J.P. Rodriguez, Phys. Rev.
 B {\bf 44}, 9582 (1991); (E)
{\bf 45}, 5119 (1992).  %; {\bf 46}, 591 (1992).

\item {6.} R. Hlubina, W.O. Putikka, T.M. Rice,
and D.V. Khveshchenko, Phys. Rev. B{\bf 46}, 11224 (1992).

\item {7.} D. Pines and P. Nozi\` eres,
{\it The Theory of Quantum Liquids}, vol. 1
(Addison-Wesley, New York, 1989).

\item {8.} H. J. Schulz, Int. J. of Mod. Phys. B{\bf 5}, 57 (1991);
M. Ogata, M.U. Luchini, S. Sorella and F.F. Assaad,
Phys. Rev. Lett. {\bf 66}, 2388 (1991).

\item {9.}  J.P. Rodriguez and B. Dou\c cot,
Phys. Rev. B{\bf 45}, 971 (1992).

\item {10.} D. Bohm and T. Staver,
Phys. Rev. {\bf 84}, 836 (1950).

\item {11.} P.M. Platzman and P.A. Wolff,
Solid State Physics (Suppl.) {\bf 13}
(Academic Press, New York, 1973).

\item {12.} V.P. Silin, Zh. Eksp. Teor. Fiz. {\bf 35}, 1243 (1958)
[Sov. Phys. JETP {\bf 8}, 870 (1959)].

\item {13.} L.P. Gorkov and I.E. Dzyaloshinskii,
Zh. Eksp. Teor. Fiz. {\bf 44}, 1650 (1963)
[Sov. Phys. JETP {\bf 17},
1111 (1963)].

\item {14.} A.A. Abrikosov, L.P. Gorkov,
and I.E. Dzyaloshinski, {\it Methods of
Quantum Field Theory in Statistical Physics},
(Dover, New York, 1975).

\item {15.} J.P. Rodriguez, unpublished.

\item {16.} Y. Suzumara, Y. Hasegawa,
and H. Fukuyama, J. Phys. Soc. Jpn. {\bf 57}, 2768 (1988).

\item {17.} N. Nagaosa and P.A. Lee,
Phys. Rev. B {\bf 45}, 966 (1992).

\item {18.} J.P. Rodriguez, Phys. Rev. B {\bf 49}, 3663 and 9831 (1994);
J.P. Rodriguez and P. Lederer,
Phys. Rev. B {\bf 48}, 16051 (1993).

\item {19.} A. Sokol and D. Pines, Phys. Rev. Lett.
{\bf 71}, 2813 (1993).

%\item {16.} See G. Shirane and R. Birgeneau, in
%{\it The Physical Properties
%of High-Temperature Superconductors},
%vol. 1, edited
%by D.M. Ginsberg (World Scientific, Singapore, 1989).

\item {20.} B. Keimer et al., Phys. Rev. B {\bf 46}, 14034 (1992).

\item {21.} P.B. Allen, Z. Fisk, and A. Migliori,
in {\it Physical Properties of High Temperature Superconductors},
edited by D.M. Ginsberg (World Scientific, Singapore, 1989).

\item {22.} S. Barisic and J. Zelanko, Europhys.
Lett. {\bf 8}, 765 (1989); R.S. Markiewicz, J. Phys.
Condensed Matter {\bf 2}, 6223 (1990).

%\item {22.} X. Zotos, P. Prelov\v sek and I. Sega,
%Phys. Rev. B {\bf 42}, 8445 (1990).

\vfill\eject
\centerline{\bf Figure Caption}
\vskip 20pt
\item {Fig. 1.}   Shown are the two branches, $c_-$ and $c_+$,
of the renormalized  phonon velocity, $c_s$,
as a function of the zero-sound velocity, $c_0$,
where the electron-phonon coupling strength is fixed at
$\rho_F g^2=0.2$ [see eq. (8a)].  The horizontal dashed line represents
bare first-sound,$^{10}$ $c_s=c_1$, while
the diagonal dashed line represents bare zero-sound, $c_s=c_0$.

\end